\documentstyle[12pt]{article}

\def\be{\begin{equation}}
\def\ee{\end{equation}}
\def\bea{\begin{eqnarray}}
\def\eea{\end{eqnarray}}

\begin{document}
\vspace*{2cm}
\begin{center}
{\large\bf ELEVEN DIMENSIONS FROM THE
MASSIVE D-2-BRANE}
\vskip 1.5 cm
{\bf Y. Lozano}\footnote{\tt Y.Lozano@fys.ruu.nl}
\vskip 0.05cm
Inst. for Theoretical Physics, University of Utrecht,\\
Princetonplein 5, 3508 TA Utrecht,\\ 
The Netherlands
\end{center}

\date{ }
\setcounter{page}{0} \pagestyle{empty}
\thispagestyle{empty}
\vskip 3 cm
\begin{abstract}
We find an eleven dimensional description of the D-2-brane
of the massive type IIA theory as a first step towards an
understanding of this theory in eleven dimensions.
By means of a world-volume IIA/M theory duality transformation 
we show that the massive D-2-brane is equivalent to the
dimensional reduction
of the eleven dimensional membrane coupled to an auxiliary 
vector field. The role of this vector field is to preserve the 
invariance under massive gauge transformations in the 
world-volume and has non-trivial dynamics, governed by a 
Chern-Simons term proportional to $1/m$. 
\end{abstract}
\vfill
\begin{flushleft}
THU-97/17  \\
hep-th/9707011\\
July 1997
\end{flushleft}
\newpage\pagestyle{plain}

\def\theequation{\thesection . \arabic{equation}}

\section{Introduction}
\setcounter{equation}{0}

It is well-known that type IIA supergravity can be 
obtained from eleven dimensional
supergravity by a dimensional reduction.
The dilaton emerges as a Kaluza-Klein field, related
to the compactification radius of the eleventh
coordinate as $R_{11}=e^{2\phi/3}$ and, therefore, in 
the strong coupling limit the theory is eleven 
dimensional. Hence
M-theory (compactified on a circle)
has been conjectured to be the strong
coupling limit of the type IIA superstring \cite{To1,W1}. 
The p-branes of type IIA supergravity have an alternative
interpretation in M-theory\footnote{This applies to all
branes but the D-8-brane, which exists for massive
type IIA \cite{BRGPT}.} \cite{DHIS,To1,To},
which for the D-branes
could be used to deduce their D=10 Lorentz covariant 
world-volume action.
In \cite{SDAST,Y,APPS1} it was shown that in order to 
obtain the complete equivalence
between the D-2 and D-4-branes of type IIA and the membrane 
\cite{BST} and 5-brane \cite{G} \cite{BRO,APPS1}
of M-theory a non-trivial transformation involving
the abelian world-volume gauge field of the D-brane had to be
made. This world-volume transformation was identified as
the underlying mapping responsible for IIA/M theory
duality \cite{revT}.

The eleven dimensional interpretation of massive
IIA supergravity (an extension of type IIA supergravity
including a cosmological constant term)
\cite{massive,BRGPT} is, however, an open problem. 
The mass parameter (proportional to the square root of
the cosmological constant) can be considered as a RR
field of the type IIA theory, since it is the dual of
the corresponding RR 9-form\footnote{The field equations
of the RR 9-form fix its dual field strength to a
constant with dimensions of mass \cite{PWPS,BRGPT}.}.
On the one hand if M-theory is
supposed to unify the different string theories 
we should be able to find an eleven dimensional 
interpretation of the massive case.
However on the other hand, there are several arguments 
showing that an extension of eleven dimensional 
supergravity to include a cosmological constant term is 
not possible \cite{BDHS}.
If such an extension exists it must be of some
unconventional form.

Our approach to study this problem will be to consider 
the D-branes of the massive type IIA theory. D-brane 
actions with $m\neq 0$
have been constructed by T-duality from the type IIB theory
\cite{BR,GHT}. The kappa symmetric extensions have been
given in \cite{BT}.

Our main result will be the derivation of an eleven 
dimensional description of
the massive D-2-brane by performing a
duality transformation on its world-volume,
in analogous way as an eleven dimensional interpretation
could be given to the D-2-brane with $m=0$
\cite{SDAST,Y}.
Now there is an explicit dependence of the world-volume action 
on the Born-Infeld field that will require a 
modification of the duality
process. The result is that the dual of the vector field 
will not be a scalar but another vector field.
We will show that, still, we can give an eleven dimensional 
interpretation in the form of an eleven dimensional membrane
compactified on a circle, coupled to an auxiliary world-volume 
vector field
whose role is to introduce invariance under massive 
gauge transformations.
This auxiliary field has non trivial dynamics, governed
by a Chern-Simons term proportional to $1/m$.

\section{The massive D-2-brane}
\setcounter{equation}{0}

The effective action of the massive D-membrane is given
by \cite{BR,GHT,BT}\footnote{We consider a Minkowskian
signature space-time.}:
\be
\label{1.1}
S=\int d^3x [e^{-\phi}\sqrt{-{\rm det}
(G_{mn}+B_{mn}+F_{mn})}+
\frac12\epsilon^{mnp}(\frac13 C_{mnp}+
C_m(F_{np}+B_{np})+\frac{m}{2}A_m\partial_n A_p)].
\ee
$(G,B)$, $(C_{(1)},C_{(3)})$ are the NS-NS, RR
fields induced in the D-brane
from ten dimensions, $F=dA$ is the field strength
associated to the BI 1-form and $m$ is the mass parameter
corresponding to the
dual field strength of the RR 9-form of type IIA.
This action is invariant under the RR transformations:
\bea
\label{1.2}
&&\delta_{RR}C_m=\partial_m\alpha\nonumber\\
&&\delta_{RR}C_{mnp}=-\partial_{[m}\alpha B_{np]}
\eea
plus the NS-NS transformations:
\bea
\label{1.3}
&&\delta C_m=-m\lambda_m\nonumber\\
&&\delta B_{mn}=-2(\partial_m\lambda_n-
\partial_n\lambda_m)\nonumber\\
&&\delta C_{mnp}=m\lambda_{[m}B_{np]}\nonumber\\
&&\delta A_m=2\lambda_m.
\eea
Notice that when $m=0$ the RR fields are invariant
under the NS-NS transformations, whereas
they transform non-trivially for $m\neq 0$.
The Chern-Simons term
$\frac{m}{4}\epsilon^{mnp}A_m\partial_n A_p$
has to be added to the action to have invariance 
under (\ref{1.3}).
This form of the WZ part was shown to be required by
T-duality from the type IIB theory \cite{BR,GHT} and also by
kappa symmetry \cite{BT}.
It can be obtained from the massless D-2-brane by redefining:
\bea
\label{1.3.1}
&&C_m\rightarrow C_m+\frac{m}{2}A_m\nonumber\\
&&C_{mnp}\rightarrow C_{mnp}-\frac{m}{2}
A_{[m}B_{np]}-\frac32 mA_m\partial_n A_p,
\eea
in such a way that the invariance under (\ref{1.3}) is
introduced\footnote{The invariant NS-NS 2-form is
$B_{mn}+\partial_m A_n-\partial_n A_m$.}. 
Quantum effects show that the mass
parameter and therefore, the cosmological constant,
has to be quantized \cite{BRGPT,GHT}.
In \cite{BRGPT} this condition is obtained by
impossing consistency between T-duality and SL(2,Z)
duality of type IIB, and in \cite{GHT} by demanding
independence of the D-brane tension on the compactification
radius.

In this paper we will focus on the bosonic part of the
action. Kappa symmetric actions have been constructed in
\cite{BT}.

When the mass is equal to zero (\ref{1.1}) is
invariant under constant translations of the abelian
gauge field. One can then construct the so-called
first order action, in which $dA$ is replaced by a
fundamental two-form whose field strength is impossed
to be zero by the introduction of a Lagrange multiplier 
(a scalar ${\tilde \Lambda}$
in a three-dimensional world-volume). Integrating out
the Lagrange multiplier the original action is obtained,
whereas the integration over the two-form yields the
dual action\footnote{An alternative way of deriving dual
actions using an auxiliary metric has been derived in
\cite{AH}.} \cite{SDAST,Y}:
\be
\label{1.4}
{\tilde S}=\int d^3x[\sqrt{-{\rm det}(e^{-2\phi/3}G_{mn}+
e^{4\phi/3}(\partial_m{\tilde \Lambda}-C_m)(\partial_n
{\tilde \Lambda}-C_n))}+\frac16\epsilon^{mnp}
(C_{mnp}+3\partial_m{\tilde \Lambda}B_{np})].
\ee
This is the Nambu-Goto action of the dimensionally reduced
eleven dimensional
supermembrane \cite{BST}, with:
\bea
\label{1.5}
&&G^{(11)}_{mn}=e^{-2\phi/3}G^{(10)}_{mn}+e^{4\phi/3}
(\partial_m{\tilde \Lambda}-C_m)(\partial_n
{\tilde \Lambda}-C_n)\nonumber\\
&&B^{(11)}_{mnp}=C^{(10)}_{mnp}+3\partial_m
{\tilde \Lambda}B_{np}.
\eea
It is invariant under the RR gauge transformations 
(\ref{1.2}) if ${\tilde \Lambda}$ transforms as 
$\delta_{RR} {\tilde \Lambda}=\alpha$, and, 
trivially, under
NS-NS transformations (up to a total derivative).

The world-volume duality transformation provides 
a way of finding the eleven dimensional (strong coupling)
description of the original action. We can then try to follow 
a similar
argument to give an eleven dimensional interpretation
to the massive IIA theory. When $m\neq 0$
the explicit dependence of the action (\ref{1.1}) on
the world-volume gauge field through the Chern-Simons term
breaks the symmetry under $A\rightarrow A+\epsilon$ needed to
obtain the dual theory in the above procedure.
However it is still possible to construct a dual action by 
building an intermediate Lagrangian from which both the initial
and dual theories can be derived.

Let us now motivate the form of this intermediate Lagrangian.
The action (\ref{1.4}) is not invariant under massive NS-NS
transformations. However we can easily modify it in such a
way that the invariance is manifest. We just need to replace
$\partial_m{\tilde \Lambda}$ by
$\partial_m{\tilde \Lambda}+{\tilde A}_m$ with
${\tilde A}_m$ transforming as
$\delta {\tilde A}_m=-m\lambda_m$, and add a Chern-Simons term:
\be
\label{1.6}
-\frac{1}{m}\epsilon^{mnp}{\tilde A}_m\partial_n{\tilde A}_p
\ee
to the WZ part.
Then the action:
\bea
\label{1.7}
{\tilde S}&=&\int d^3x [\sqrt{-{\rm det}(e^{-2\phi/3}G_{mn}
+e^{4\phi/3}(\partial_m{\tilde \Lambda}+{\tilde A}_m-C_m)
(\partial_n{\tilde \Lambda}+{\tilde A}_n-C_n))}+\nonumber\\
&&\frac16 \epsilon^{mnp}(C_{mnp}+
3(\partial_m{\tilde \Lambda}+{\tilde A}_m) B_{np})-
\frac{1}{m}\epsilon^{mnp}{\tilde A}_m
\partial_n{\tilde A}_p]
\eea
is invariant under the two transformations (\ref{1.2}) 
and (\ref{1.3}) plus $\delta_{RR}{\tilde \Lambda}=\alpha$
and $\delta {\tilde A}_m=-m\lambda_m$.
Notice that ${\tilde A}_m$ can be taken as a coexact 
1-form, since the NS-NS transformations (\ref{1.3}) are 
non-trivial only if $\lambda$ is coexact. If $\lambda$ is
exact\footnote{We consider topologically trivial 
world-volumes.} the NS-NS backgrounds are 
invariant\footnote{$A_m$ transforms with a total derivative
but it doesn't contribute to the action.} and the RR fields
transform as (\ref{1.2}) with $\alpha$ depending on the
mass, and this transformation is already cancelled by 
$\delta_{RR}{\tilde \Lambda}=\alpha$.
We can then work with a fundamental 1-form 
${\tilde V}_m\equiv\partial_m{\tilde \Lambda}+
{\tilde A}_m$ whose exact component is given by the 
differential of the eleventh coordinate and the coexact
one by the auxiliary field ${\tilde A}_m$.

We now show that in fact (\ref{1.7}) is equivalent to the
massive D-2-brane action under a world-volume duality 
transformation.
We just need to realize that it can be obtained from the
intermediate action:
\bea
\label{1.8}
S_I&=&\int d^3x [\sqrt{-{\rm det}(e^{-2\phi/3}G_{mn}+
e^{4\phi/3}({\tilde V}_m-C_m)({\tilde V}_n-C_n))}+
\frac16\epsilon^{mnp}C_{mnp}+
\frac12\epsilon^{mnp}{\tilde V}_m B_{np}+\nonumber\\
&&\frac{m}{4}\epsilon^{mnp}A_m\partial_n A_p+\frac12
\epsilon^{mnp}{\tilde V}_m(\partial_nA_p-\partial_pA_n)]
\eea
after integration over the auxiliary field $A_m$. 
The equation of motion for $A_m$ gives the constraint
$\epsilon^{mnp}(\partial_n{\tilde V}_p+\frac{m}{2}
\partial_nA_p)=0$, which substituted in (\ref{1.8}) gives
the action (\ref{1.7}) up to a total derivative.

If instead we integrate out ${\tilde V}_m$ it is easy
to check that the action for the massive D-2-brane is 
obtained.
For this purpose it is helpful to write:
\be
\label{1.9}
\sqrt{-{\rm det}(e^{-2\phi/3}G_{mn}+
e^{4\phi/3}({\tilde V}_m-C_m)({\tilde V}_n-C_n))}=
\sqrt{-{\rm det}g}\sqrt{1+e^{4\phi/3}
({\tilde V}_m-C_m)({\tilde V}^m-C^m)}
\ee
where $g_{mn}=e^{-2\phi/3}G_{mn}$.
Therefore when $m\neq 0$ the dual of the vector field $A_m$
is also a vector field\footnote{This is a well-known result in 
topologically massive three dimensional gauge theories (see
for instance \cite{DJ}). Our intermediate action (\ref{1.8})
is the generalization to the Born-Infeld Lagrangian of the 
intermediate action presented there.} ${\tilde V}_m$, 
that can be decomposed into an exact 
component, playing the role of the (differential of the)
eleventh coordinate, plus a coexact component.
(\ref{1.7}) then provides
an eleven dimensional description
of the massive D-2-brane as
the action of the eleven dimensional membrane compactified
on a circle, coupled to an auxiliary world-volume field
required to introduce invariance under massive gauge 
transformations. This field
has non-trivial dynamics, dictated by the Chern-Simons term
proportional to $1/m$. As a check, when the mass is sent to 
zero the Chern-Simons term in (\ref{1.7}) vanishes, implying
that ${\tilde A}_m=0$ (since it is a coexact 1-form),
and the action for the dimensional
reduction of the eleven dimensional membrane (\ref{1.4})
is recovered. 

It is worth noting that (\ref{1.7}) can be obtained from
the massless dual action (\ref{1.4}) after the redefinitions:
\bea
\label{1.10}
&&C_m\rightarrow C_m-{\tilde A}_m\nonumber\\
&&B_{mn}\rightarrow B_{mn}-\frac{2}{m}
(\partial_m{\tilde A}_n-\partial_n{\tilde A}_m)\nonumber\\
&&C_{mnp}\rightarrow C_{mnp}+{\tilde A}_{[m}B_{np]}
-\frac{6}{m}{\tilde A}_m\partial_n{\tilde A}_p
\eea
are made. 
Comparing these expressions with (\ref{1.3.1}) we see
that ${\tilde A}$ plays the same role than the original
BI field $A$ (note the different scalings with the mass).
In both the original and dual theories these fields need
to be introduced in order to have invariance under the
NS-NS transformations (\ref{1.3}).

We give now some speculations about the possible space-time
interpretation of the auxiliary field ${\tilde A}$.
The redefinitions (\ref{1.3.1}) and (\ref{1.10}) 
resemble the transformations of
the ten dimensional background fields: 
\bea
\label{1.11}
&&B^\prime=B-\frac{2}{m}dC_{(1)}\nonumber\\
&&C^\prime_{(3)}=C_{(3)}+3C_{(1)}B-
\frac{3}{m}C_{(1)}dC_{(1)},
\eea
required to formulate Romans' massive IIA supergravity 
\cite{massive} in a way that the invariance under the
gauge symmetries:
\bea
\label{1.12}
&&\delta C_{(1)}=-m\lambda\nonumber\\
&&\delta B=-2d\lambda\nonumber\\
&&\delta C_{(3)}=3m\lambda B
\eea
is manifest. The RR 1-form is then said to play the role
of a Stueckelberg field, introducing the symmetry
under massive gauge transformations.
The NS-NS 2-form acquires mass 
by absorbing the RR 1-form, in a Higgs-type mechanism
(see \cite{massive,BRGPT} for details).
The space-time transformations (\ref{1.11}) are obtained
in the world-volume if the induced RR 1-form
absorbs the auxiliary vector field $A$ or ${\tilde A}$
(depending on whether we are in the original or dual theories).
Perhaps this could give a hint on the possible space-time
interpretation of this field.

The obvious next step would be to find the eleven dimensional
supergravity extension containing (\ref{1.7}) as a solution.
An idea for this purpose is to find the
kappa symmetric extension of this action, since for $m=0$
it is known
that kappa symmetry of the
eleven dimensional membrane implies that the backgrounds must
satisfy the field equations of 
eleven dimensional supergravity \cite{BST}.
Along these lines it has been shown in \cite{BCT} that the 
field equations of massive IIA supergravity are implied
by kappa symmetry of the massive D-2-brane.
We leave this point for further investigation.

\section{Conclusions}
\setcounter{equation}{0}

We have seen that the massive D-membrane of the type IIA 
theory can be
interpreted as the eleven dimensional membrane 
compactified on a circle, coupled to an auxiliary world-volume
vector field needed to introduce invariance under the 
massive gauge
transformations induced in the world-volume.
This field has non-trivial
dynamics, governed by a Chern-Simons term proportional to
$1/m$. 
Its physical meaning remains an open
problem which we hope to address in a near future.

A similar type of effective action induced from 
eleven dimensions with an additional
vector field has been presented in \cite{BJO} in the
context of the eleven dimensional Kaluza-Klein monopole. 
This solution has eight translational isometries but 
only seven can be interpreted as 
world-volume directions \cite{H}.
This implies that there is an extra scalar in the
world-volume action, that needs to
be removed in order to have the correct degrees of freedom.
The mechanism proposed in \cite{BJO} is to gauge the isometry 
under translations
on the eighth direction by introducing a vector field in
the world-volume action.
It could be interesting
to analyze if a similar type of interpretation could be
applicable to our effective action.

The equivalence between the original and dual actions 
that we have presented relies on a saddle-point
approximation. This was also the case for
the D-2-brane/eleven dimensional membrane
duality of \cite{SDAST}. In that case it was possible to
show the equivalence of the two theories
at the level of their partition
functions by formulating the duality as
a canonical transformation in the phase space associated to
the BI field of the D-2-brane \cite{Y}.  
It is yet unclear to us whether this is also the case in the
present theories.

Finally, it would be interesting to find similar eleven
dimensional descriptions for other
D-branes of massive type IIA, and in particular for the 
D-8-brane. This could shed some light on the determination
of the action of the eleven dimensional 9-brane
\cite{BRGPT,BREJS,H}.

\subsection*{Acknowledgements}

I would like to thank E. Alvarez, B. de Wit, J. Gomis,
M.B. Green and C. Schmidhuber for very useful discussions.
Work supported by the European Commission TMR
programme ERBFMRX-CT96-0045.

\end{document}